\documentclass[twocolumn, showpacs, preprintnumbers, superscriptaddress, aps, prl, floatfix, tightenlines]{revtex4-1}         
\pdfoutput=1
\usepackage{hyperref}
\usepackage{amsmath}
\usepackage{cleveref}
\usepackage{CJKutf8}
\usepackage{latexsym}
\usepackage{amstext}
\usepackage{amsfonts}
\usepackage{dsfont}
\usepackage{color}
\usepackage{amssymb}
\usepackage{relsize}
\usepackage{amsxtra}
\usepackage{verbatim}
\usepackage{hyperref}
\usepackage{graphics}
\usepackage{graphicx}
\usepackage{multirow}

\AtBeginDvi{\input{zhwinfonts}}
\renewcommand{\thetable}{\arabic{table}}

\begin{document}
\begin{CJK*}{UTF8}{shiai}

\title{Crystalline and magnetic structures, magnetization, heat capacity, and anisotropic magnetostriction effect in a yttrium-chromium oxide}

\author{Yinghao Zhu} 
\thanks{These three authors contributed equally.}
\affiliation{Joint Key Laboratory of the Ministry of Education, Institute of Applied Physics and Materials Engineering, University of Macau, Avenida da Universidade, Taipa, Macao SAR 999078, China}
\author{Ying Fu} 
\thanks{These three authors contributed equally.}
\affiliation{Joint Key Laboratory of the Ministry of Education, Institute of Applied Physics and Materials Engineering, University of Macau, Avenida da Universidade, Taipa, Macao SAR 999078, China}
\author{Bao Tu} 
\thanks{These three authors contributed equally.}
\affiliation{Joint Key Laboratory of the Ministry of Education, Institute of Applied Physics and Materials Engineering, University of Macau, Avenida da Universidade, Taipa, Macao SAR 999078, China}
\affiliation{Department of Materials Science and Engineering, Shenzhen Key Laboratory of Full Spectral Solar Electricity Generation (FSSEG), Southern University of Science and Technology, No. 1088, Xueyuan Rd., Shenzhen 518055, Guangdong, China}
\author{Tao Li} 
\affiliation{Neutron Scattering Technical Engineering Research Center, School of Mechanical Engineering, Dongguan University of Technology, Dongguan 523808, China}
\author{Jun Miao} 
\affiliation{School of Materials Science and Engineering, University of Science and Technology Beijing, Beijing 100083, China}
\author{Qian Zhao} 
\affiliation{Joint Key Laboratory of the Ministry of Education, Institute of Applied Physics and Materials Engineering, University of Macau, Avenida da Universidade, Taipa, Macao SAR 999078, China}
\author{Si Wu} 
\affiliation{Joint Key Laboratory of the Ministry of Education, Institute of Applied Physics and Materials Engineering, University of Macau, Avenida da Universidade, Taipa, Macao SAR 999078, China}
\author{Junchao Xia} 
\affiliation{Joint Key Laboratory of the Ministry of Education, Institute of Applied Physics and Materials Engineering, University of Macau, Avenida da Universidade, Taipa, Macao SAR 999078, China}
\author{Pengfei Zhou} 
\affiliation{Joint Key Laboratory of the Ministry of Education, Institute of Applied Physics and Materials Engineering, University of Macau, Avenida da Universidade, Taipa, Macao SAR 999078, China}
\author{Ashfia Huq} 
\affiliation{Oak Ridge National Laboratory, Oak Ridge, Tennessee 37831, USA}
\author{Wolfgang Schmidt} 
\affiliation{Forschungszentrum J$\ddot{u}$lich GmbH, J$\ddot{u}$lich Centre for Neutron Science at ILL, 71 avenue des Martyrs, Grenoble 38042, France}
\author{Defang Ouyang} 
\affiliation{State Key Laboratory of Quality Research in Chinese Medicine, Institute of Chinese Medical Sciences (ICMS), University of Macau, Avenida da Universidade, Taipa, Macao SAR 999078, China}
\author{Defang Ouyang} 
\affiliation{State Key Laboratory of Quality Research in Chinese Medicine, Institute of Chinese Medical Sciences (ICMS), University of Macau, Avenida da Universidade, Taipa, Macao SAR 999078, China}
\author{Zikang Tang} 
\email{zktang@um.edu.mo}
\affiliation{Joint Key Laboratory of the Ministry of Education, Institute of Applied Physics and Materials Engineering, University of Macau, Avenida da Universidade, Taipa, Macao SAR 999078, China}
\author{Zhubing He} 
\email{hezb@sustc.edu.cn}
\affiliation{Department of Materials Science and Engineering, Shenzhen Key Laboratory of Full Spectral Solar Electricity Generation (FSSEG), Southern University of Science and Technology, No. 1088, Xueyuan Rd., Shenzhen 518055, Guangdong, China}
\author{Hai-Feng Li} 
\email{haifengli@um.edu.mo}
\affiliation{Joint Key Laboratory of the Ministry of Education, Institute of Applied Physics and Materials Engineering, University of Macau, Avenida da Universidade, Taipa, Macao SAR 999078, China}

\date{\today}

\begin{abstract}

We have studied a nearly stoichiometric insulating Y$_{0.97(2)}$Cr$_{0.98(2)}$O$_{3.00(2)}$ single crystal by performing measurements of magnetization, heat capacity, and neutron diffraction. Albeit that the YCrO$_3$ compound behaves like a soft ferromagnet with a coersive force of $\sim$0.05 T, there exist strong antiferromagnetic (AFM) interactions between Cr$^{3+}$ spins due to a strongly negative paramagnetic Curie-Weiss temperature, i.e., --433.2(6) K. The coexistence of ferromagnetism and antiferromagnetism may indicate a canted AFM structure. The AFM phase transition occurs at $T_\textrm{N} =$ 141.5(1) K, which increases to $T_\textrm{N}$(5T) = 144.5(1) K at 5 T. Within the accuracy of the present neutron-diffraction studies, we determined a G-type AFM structure with a propagation vector \textbf{k} = (1 1 0) and Cr$^{3+}$ spin directions along the crystallographic \emph{c} axis of the orthorhombic structure with space group \emph{Pnma} below $T_\textrm{N}$. At 12 K, the refined moment size is 2.45(6) $\mu_\textrm{B}$, $\sim$82\% of the theoretical saturation value 3 $\mu_\textrm{B}$. The Cr$^{3+}$ spin interactions are probably two-dimensional Ising like within the reciprocal (1 1 0) scattering plane. Below $T_\textrm{N}$, the lattice configuration (\emph{a}, \emph{b}, \emph{c}, and \emph{V}) deviates largely downward from the Gr$\ddot{\textrm{u}}$neisen law, displaying an anisotropic magnetostriction effect and a magnetoelastic effect. Especially, the sample contraction upon cooling is enhanced below the AFM transition temperature. There is evidence to suggest that the actual crystalline symmetry of YCrO$_3$ compound is probably lower than the currently assumed one. Additionally, we compared the $t_{2\textrm{g}}$ YCrO$_3$ and the $e_\textrm{g}$ La$_{7/8}$Sr$_{1/8}$MnO$_3$ single crystals for a further understanding of the reason for the possible symmetry lowering.

\end{abstract}

\maketitle
\end{CJK*}


\section{I. Introduction}

Magnetic materials with ferroelectricity are very interesting because a spontaneous electric polarization exists within them to give the substance extraordinary physical and electronic properties and a wide variety of applications such as data storage, catalysts, fuel cells, and sensors \cite{Scott1998, Vanaken2004, Zhao2014}. The magnetic and ferroelectric properties can be coupled with each other, therefore, an applied magnetic/electric field is able to switch the electric polarization/magnetization. Above the ferroelectric phase transition, the ferroelectric materials are usually centrosymmetric structurally and behave as an ordinary dielectric; below the phase transition, an electrically-polarized phase forms spontaneously with a noncentrosymmetric structure in the weakly-coupled systems \cite{Gibbs2011} or can be induced by a magnetic phase transition in the strongly-coupled families \cite{Kenzelmann2005, Mostovoy2006}. Therefore, to build the actual structural and magnetic models may shed light on the nature of the ferroelectric phase transition.

\begin{figure}[!t]
\centering
\includegraphics[width = 0.48\textwidth] {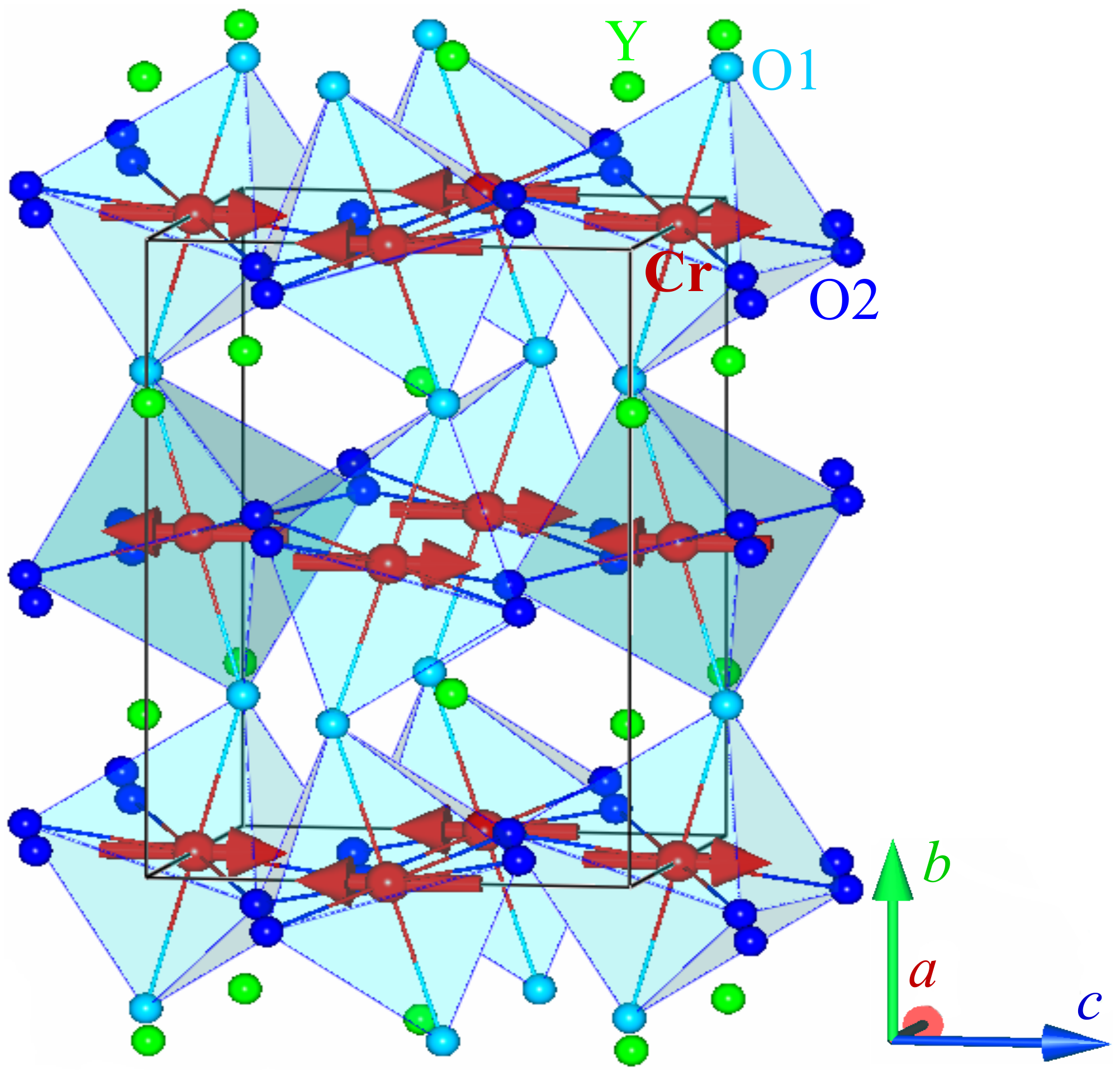}
\caption{
Orthorhombic crystal structure (with space group $Pnma$) with one unit cell (solid lines) and the AFM structure in one AFM unit cell with the propagation vector at \textbf{k} = (1 1 0) below $T_\textrm{N}$ = 141.5(1) K of the YCrO$_3$ single crystal. The arrows on the Cr ions represent the spins of chromium. Both the unit cells of orthorhombic and AFM structures are (\emph{a b c}).
}
\label{Figure1}
\end{figure}

Most ferroelectric materials are perovskite-based oxides. In 1954, Looby and Katz replaced lanthanum in LaCrO$_3$ compound with yttrium during searching for new perovskite-type families and synthesized the YCrO$_3$ compound with an impurity of $\sim$2.5\% Cr$_2$O$_3$ using NaCl as the flux under a hydrogen atmosphere \cite{Looby1954}. Based on the observation of a very weak extra Bragg peak, they indexed the x-ray powder-diffraction pattern with a monoclinic cell ($a = c =$ 7.61 {\AA}, and $b =$ 7.54 {\AA}) by doubling the fundamental perovskite unit cell \cite{Looby1954}. One year later, the crystal structure (Fig.~\ref{Figure1}) was determined to be orthorhombic (with space group $Pbnm$) with unit-cell constants $a = 5.238$, $b = 5.518$, and $c = 7.54$ {\AA} \cite{Geller1956}. This structure becomes more distorted as the pressure increases \cite{Ardit2010}. The lanthanide orthochromites of general formula RECrO$_3$ (RE = rare earth and Y) can be prepared by four different self-propagating high-temperature syntheses \cite{Kuznetsov1998}, i.e., the amorphous citrate precursor method \cite{Poplawski2000}, the conventional solid-state reaction method \cite{Westphal2000, Li2008, Zhu2019-2}, the hydrothermal synthesis \cite{Sardar2011}, and the microwave-assisted technique \cite{Gonjal2013}. Chemical substitution effect in the Y$_{1-\textrm{x}}$M$_\textrm{x}$CrO$_3$ (M = Mg, Ca, Sr, Ba) compound \cite{Weber1986} and defect chemistry of the Ca-doped YCrO$_3$ compound \cite{Carini1991} were investigated. The YCrO$_3$ compound in forms of bulk and thin film was suggested to be a candidate material for high-temperature thermistors \cite{Kagawa1997, Kim2003, Seo2015}. Among the catalysts of ABO$_3$-type perovskite oxides (A = La, Y, Nd, Gd; B = Fe, Mn, Cr, Co) for the oxidation of 1,2-dichlorobenzene, the YCrO$_3$ compound was found to be the most active catalyst and was the only one that displayed no loss of its initial activity after several hours on stream \cite{Poplawski2000}. The studies of nanocrystalline (un)doped YCrO$_3$ materials were reported \cite{Bedekar2007, Bahadur2008, Sinha2019}. The Nd-doped YCrO$_3$ nanoparticles display a semiconducting feature and an enhanced dc conductivity as the  Nd content increases, following the Dyre's free energy barrier model \cite{Sinha2019}. The magnetic configuration of the YCrO$_3$ compound below $T_\textrm{N} =$ 140 K was proposed to be a canted antiferromagnetic (AFM) structure with antisymmetric spin superexchanges \cite{Judin1966, Bertaut1966, Morishita1981}. It was reported that there existed a spin reorientation of the Cr$^{3+}$ moments in the YCrO$_3$ compound at $\sim$60 K, corresponding to a rotation of the AFM easy axis \cite{Duran2010}. Ferrimagnetism was found in the half-doped YMn$_{0.5}$Cr$_{0.5}$O$_3$ compound \cite{Hao2014}. The magnetic interactions in bulk YCrO$_3$ compound were classified as classical three-dimensional isotropic Heisenberg universality according to the $ab$ $initio$ calculations and Monte Carlo simulations based on a cubic structure with space group $Pm3m$ and lattice constant $a =$ 3.76 {\AA} \cite{Ahmed2019}. To understand the ferroelectric anomaly occurring at $\sim$473 K, the first-principles density functional theory calculations found that the noncentrosymmetric monoclinic structure (with space group $P2_1$) was the stablest one in view of its lowest energy \cite{Serrao2005}. The YCrO$_3$ compound was reported to be a relaxor ferroelectric material at about 450 K because of the local noncentrosymmetric structure \cite{Duran2010}. The high-temperature magnetism (300--980 K) and crystallographic information (321--1200 K) were studied by a time-of-flight neutron powder diffraction \cite{Zhu2019-2}, and it is of great interest that the structural information such as lattice constants, space group, bond angles, bond lengths, and the local distortion parameter have no response to the dielectric anomaly observed around 473 K \cite{Serrao2005}. There has been a long-standing debate about the decision as to which structural symmetry is correct \cite{Looby1954, Geller1956, Ardit2010, Kuznetsov1998, Poplawski2000, Westphal2000, Li2008, Zhu2019-2, Sardar2011, Gonjal2013, Weber1986, Carini1991, Kagawa1997, Kim2003, Seo2015, Bedekar2007, Bahadur2008, Sinha2019, Judin1966, Bertaut1966, Morishita1981, Hao2014, Ahmed2019, Serrao2005, Duran2010, Katz1955, Ziel1969, Remeika1956, Grodkiewicz1966, Todorov2011, Sugano1971, Zhu2019}, which necessitates a growth of the high-quality YCrO$_3$ single crystal \cite{Remeika1956, Grodkiewicz1966, Todorov2011, Sugano1971, Zhu2019} and a careful study of its crystalline and magnetic properties. Previously, small yttrium chromite single crystals with millimeter in size were grown from the PbF$_2$-B$_2$O$_3$ or the PbF$_2$-B$_2$O$_3$-KF flux in a platinum crucible \cite{Remeika1956, Grodkiewicz1966, Todorov2011}.

\begin{figure*}[!t]
\centering \includegraphics[width = 0.88\textwidth] {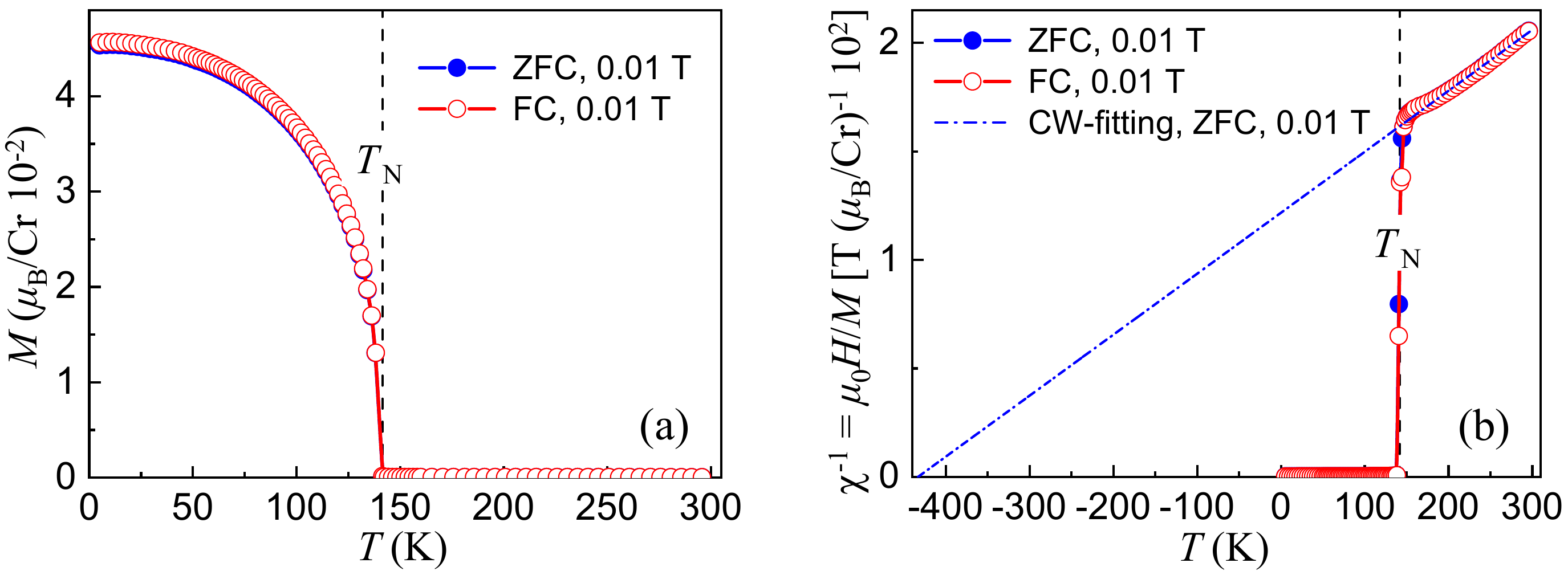}
\caption{
(a) ZFC and FC magnetization (\emph{M}) of chromium ions in the single-crystal YCrO$_3$ compound as a function of temperature measured at $\mu_0H = $ 0.01 T.
(b) Corresponding ZFC and FC inverse magnetic susceptibility $\chi^{-1}$ (circles) of chromium ions in the single-crystal YCrO$_3$ compound versus temperature. The dash-dotted line indicates a CW behavior of the ZFC data at elevated temperatures between 200 and 300 K, which was extrapolated to $\chi^{-1} = 0$ to show the PM Curie temperature $\theta_{\textrm{CW}}$. The fit results were listed in Table~\ref{Table1}. In (a) and (b), $T_\textrm{N} = 141.5$(1) K labels the AFM transition temperature at $\mu_0H = $ 0.01 T, and the solid lines are guides to the eye.
}
\label{Figure2}
\end{figure*}

\begin{figure}[!t]
\centering \includegraphics[width = 0.48\textwidth] {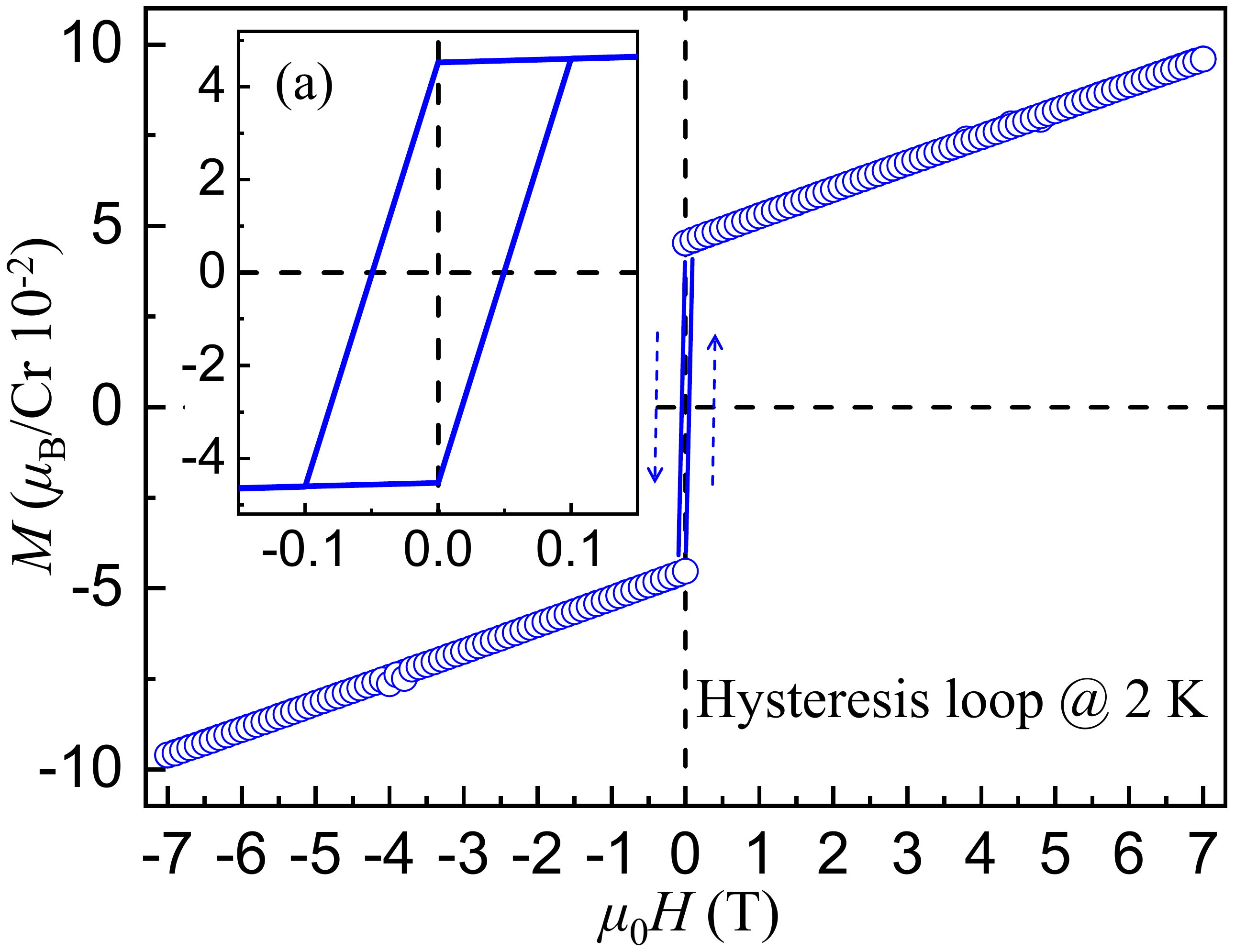}
\caption{
ZFC magnetic hysteresis loop of the single-crystal YCrO$_3$ compound measured at 2 K. Inset (a) is the enlarged image of the narrow loop.
}
\label{Figure3}
\end{figure}

\begin{table}[!t]
\renewcommand*{\thetable}{\Roman{table}}
\caption{Theoretical quantum numbers of YCrO$_3$ compound: spin \emph{S}, orbital \emph{L}, total angular momentum \emph{J}, as well as the ground-state term $^{2S+1}L_J$. Due to a quenching by the hosted crystal field, the actual orbital angular momentum \emph{L} = 0 for the 3\emph{d} ions in most cases, leading to the Land$\acute{\textrm{e}}$ factor $g_J$ = 2. We also summarized the theoretical (theo.) and measured (meas.) [Fig. 2(b)] values of effective (eff) chromium moment $\mu_{\textrm{eff}}$, PM Curie temperature $\theta_{\textrm{CW}}$, theoretical saturation (sat) chromium moment $\mu_{\textrm{sat{\_}theo.}}$, and AFM transition N$\acute{\textrm{e}}$el temperatures ($T_{\textrm{N}}$) at $\mu_0H$ = 0.01 and 5 T. The refined chromium moment size ($\mu_{\textrm{meas.}}$) at 12 K with the AFM model as shown in Fig. 1 from our POWGEN study was listed. The numbers in parenthesis are the estimated standard deviations of the last significant digit.}
\label{Table1}
\begin{ruledtabular}
\begin{tabular} {lcl}
\multicolumn{3}{c} {A YCrO$_3$ single crystal}                                                                              \\*
\hline
3$d$ ion                                                                         &          & Cr$^{3+}$                     \\*
3$d^\textrm{n}$                                                                  &          & 3                             \\*
$S$                                                                              &          & 3/2                           \\*
$L$                                                                              &          & 3                             \\*
$J = L - S$ (Hund{\textquoteright}s rule for free Cr$^{3+}$)                     &          & 3/2                           \\*
$^{2S+1}L_J$                                                                     &          & $^4F_{\frac{3}{2}}$           \\*
\hline
$g_J$ (quenched $L$ = 0, $J$ = $S$)                                              &          & 2                             \\*
$\mu_{\textrm{eff{\_}theo.}} = g_J [S(S + 1)]^{\frac{1}{2}}$ $(\mu_\textrm{B})$  &          & 3.873                         \\*
$\mu_{\textrm{sat{\_}theo.}} = g_J S$ $(\mu_\textrm{B})$                         &          & 3                             \\*
\hline
$\mu_{\textrm{eff{\_}meas.}}$ $(\mu_\textrm{B})$                                 &          & 3.95(2)                       \\*
$\theta_{\textrm{CW}}$ (K)                                                       &          & --433.2(6)                    \\*
$T_\textrm{N}$ (at 0.01 T)                                                       &          & 141.5(1)                      \\*
$T_\textrm{N}$ (at 5 T)                                                          &          & 144.5(1)                      \\*
$\mu_{\textrm{meas.}}$ (12 K, POWGEN) $(\mu_\textrm{B})$                         &          & 2.45(6)                       \\*
\end{tabular}
\end{ruledtabular}
\end{table}

In this paper, we have synthesized a centimeter-sized YCrO$_3$ single crystal with a laser diode floating-zone (FZ) furnace \cite{Zhu2019} and performed measurements of the chemical compositions, resistivity, and magnetization as functions of temperature and applied-magnetic field, heat capacity, time-of-flight neutron-powder diffraction based on a spallation neutron source, and single-crystal neutron diffraction based on a reactor. The chemical compositions of the grown YCrO$_3$ single crystal are nearly stoichiometric, and the YCrO$_3$ compound is a robust insulator.

\begin{figure}[!t]
\centering \includegraphics[width = 0.48\textwidth] {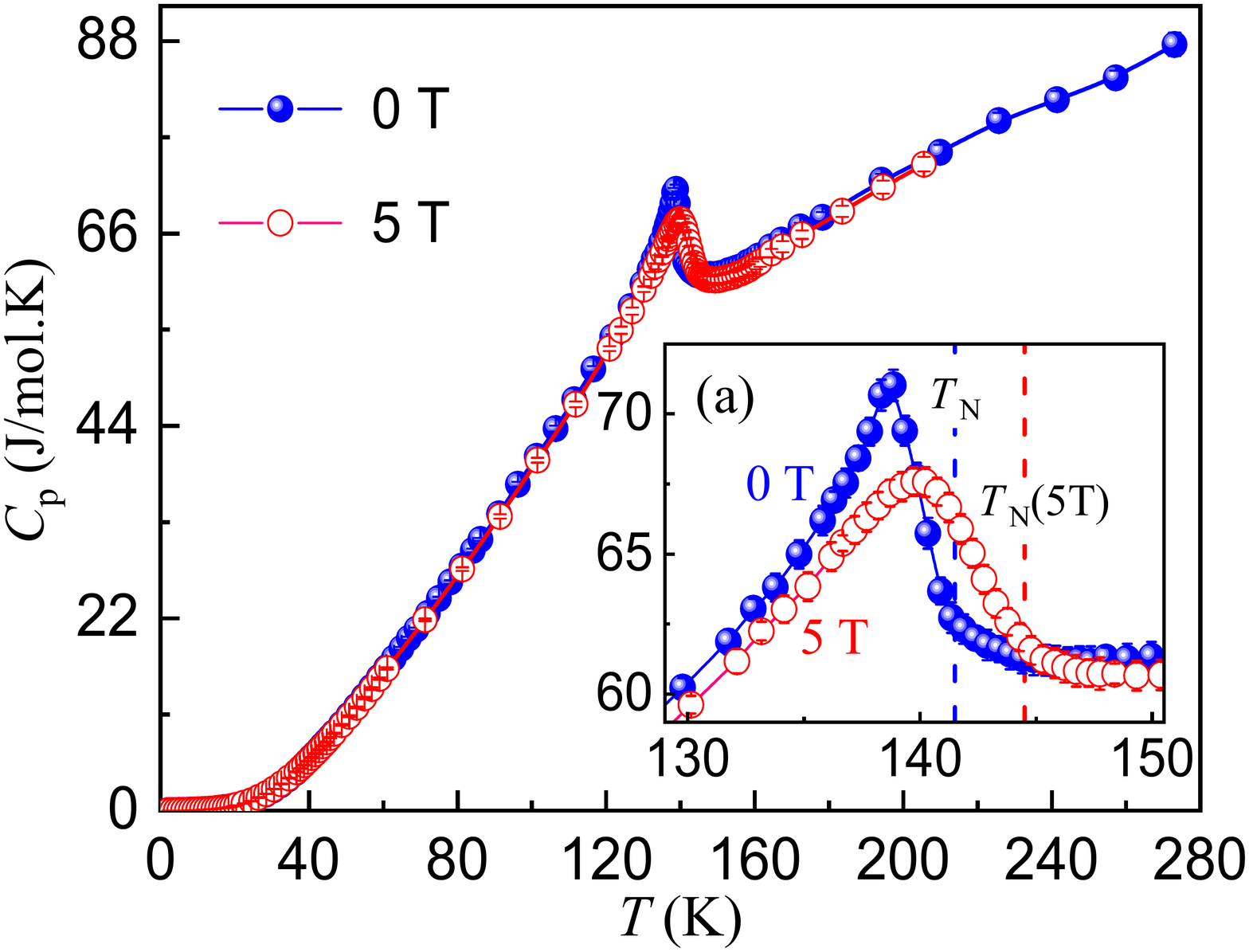}
\caption{
Heat capacities of the single-crystal YCrO$_3$ compound measured at 0 T (solid circles) and 5 T (void circles). The solid lines are guides to the eye. Inset (a) is the enlarged image around the AFM transition temperatures. The vertical dashed lines show the detailed transition temperatures at the fields of 0 and 5 T. Here, $T_\textrm{N}$(0T) = 141.5(1) K at 0 T; by comparison, at 5 T, $T_\textrm{N}$(5T) = 144.5(1) K. The solid lines are guides to the eye.
}
\label{Figure4}
\end{figure}

\section{II. Experimental}

The preparation of polycrystalline samples of YCrO$_3$ compound has been described previously in detail \cite{Zhu2019-2}. The single crystals of YCrO$_3$ compound were grown by the FZ method \cite{Li2008} with a laser diode FZ furnace (Model: LD-FZ-5-200W-VPO-PC-UM) \cite{Zhu2019, Wu2020}. We employed inductively coupled plasma with optical emission spectroscopy (ICP-OES) analysis to quantitatively determine chemical compositions of the investigated single crystals.

We measured resistivity, magnetization, and heat capacity of the YCrO$_3$ samples with a Quantum Design physical property measurement system (PPMS DynaCool instrument). Resistivity measurements were carried out with the standard four-probe method at zero field from 2 to 300 K. The dc magnetization measurements at an applied-magnetic field of 0.01 T were carried out with two modes from 5 to 295 K: One was after cooling with zero magnetic field (ZFC), and the other was under the applied-magnetic field (FC). Magnetic hysteresis loop from 7 to --7 T and then back to 7 T was measured at 2 K. Heat capacities were measured at 0 (2--273 K) and 5 T (2--205 K).

We pulverized one grown YCrO$_3$ single crystal ($\sim$4 g) with a Vibratory Micro Mill (FRITSCH PULVERISETTE 0) and performed a time-of-flight neutron-powder diffraction study on the POWGEN diffractometer (SNS, USA) from 12 to 302 K at 0 T. The \emph{d} band covers a range of 0.78--7.77 \AA. Neutron holds a magnetic moment, thus, neutron scattering is a powerful technique for solving magnetic structures \cite{Li2009, Xiao2010, Zhang2013, Xiao2019}. The higher \emph{d} band (1.7--7.77 \AA) is able to monitor all magnetic Bragg reflections that are used to refine the low-temperature magnetic structure. Data from the lower \emph{d} band (0.78--3.00 \AA) give indication of all possible structural phase transitions. We analyzed all collected time-of-flight neutron-powder diffraction data with the software of FULLPROF SUITE \cite{Rodriguez-Carvajal1993}. We refined scale factor, lattice constants, zero epithermal shift, background, peak profile shape, atomic positions, isotropic thermal parameters, and preferred orientation.

Single-crystal neutron diffraction was performed at the D23 diffractometer, located at the Institut Laue–Langevin (ILL), France.

\section{III. Results and discussion}

\begin{figure*}[!t]
\centering \includegraphics[width = 0.78\textwidth] {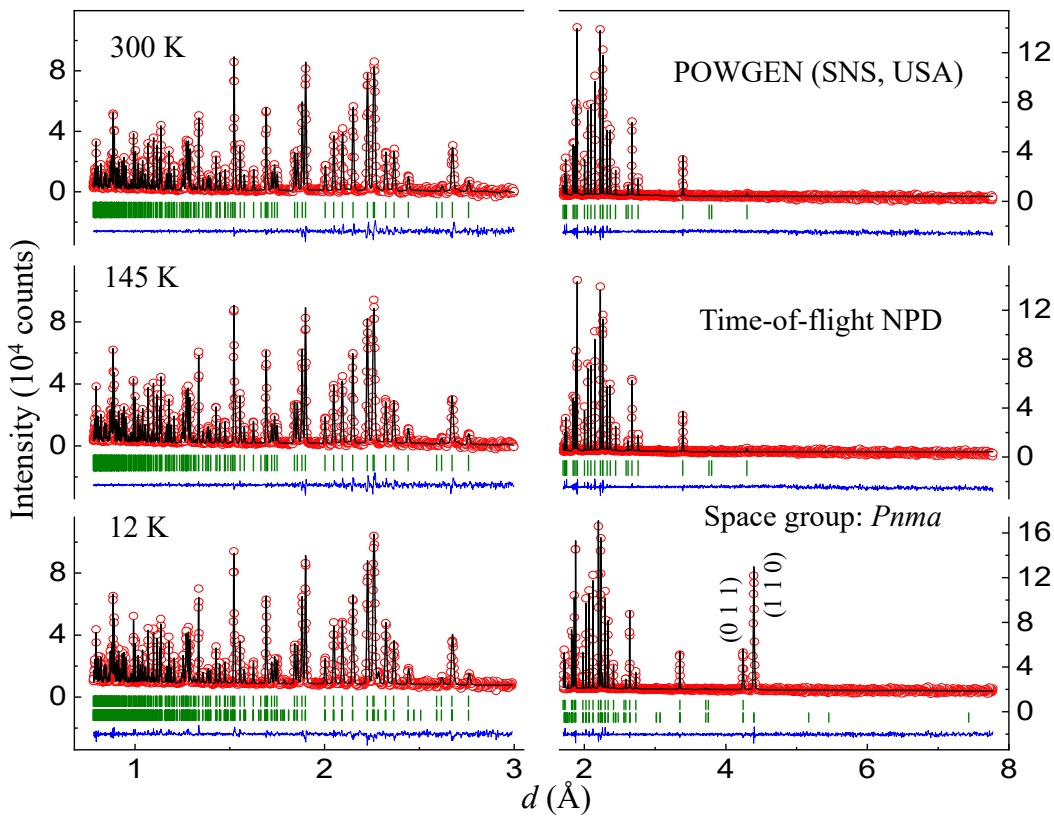}
\caption{
Observed (circles) and calculated (solid lines) time-of-flight neutron-powder diffraction (NPD) patterns of a pulverized YCrO$_3$ single crystal, collected from the POWGEN diffractometer (SNS, USA) at 12, 145, and 300 K. The vertical bars mark the positions of nuclear (up, space group $Pnma$) and magnetic (down, space group $P$-1) Bragg reflections, and the lower curves represent the difference between observed and calculated patterns.
}
\label{Figure5}
\end{figure*}

\begin{figure}[!t]
\centering \includegraphics[width = 0.48\textwidth] {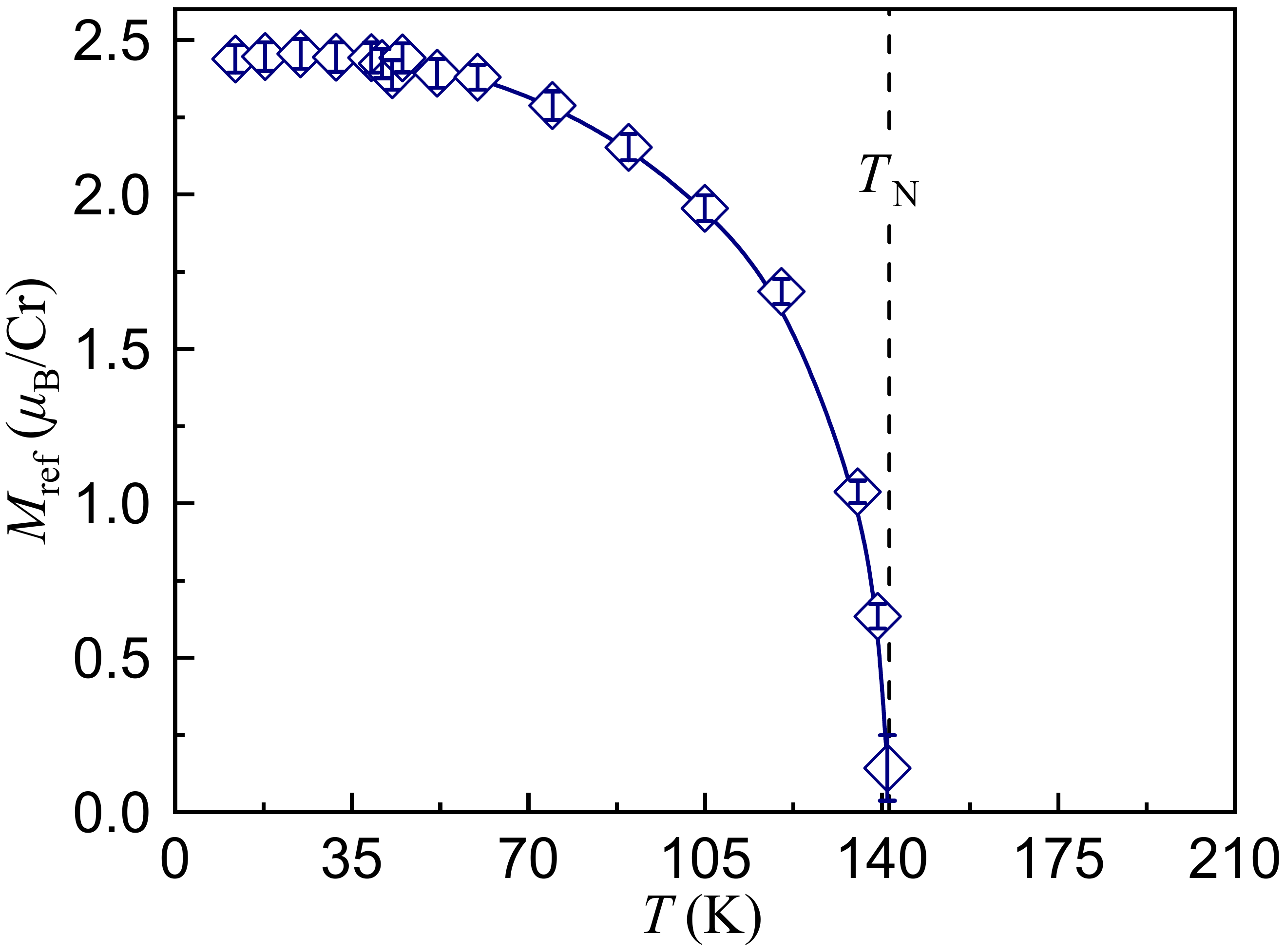}
\caption{
Refined chromium-moment size $M_{\textrm{ref}}$ (at 0 T) of a pulverized YCrO$_3$ single crystal versus temperature by the software of FULLPROF SUITE \cite{Rodriguez-Carvajal1993}. The solid line is a guide to the eye. Error bars are standard deviations obtained from our FULLPROF refinements in the \emph{Pnma} symmetry. $T_\textrm{N}$ = 141.5(1) K labels the AFM transition temperature at zero applied-magnetic field.
}
\label{Figure6}
\end{figure}

\begin{figure*}[!t]
\centering \includegraphics[width = 0.88\textwidth] {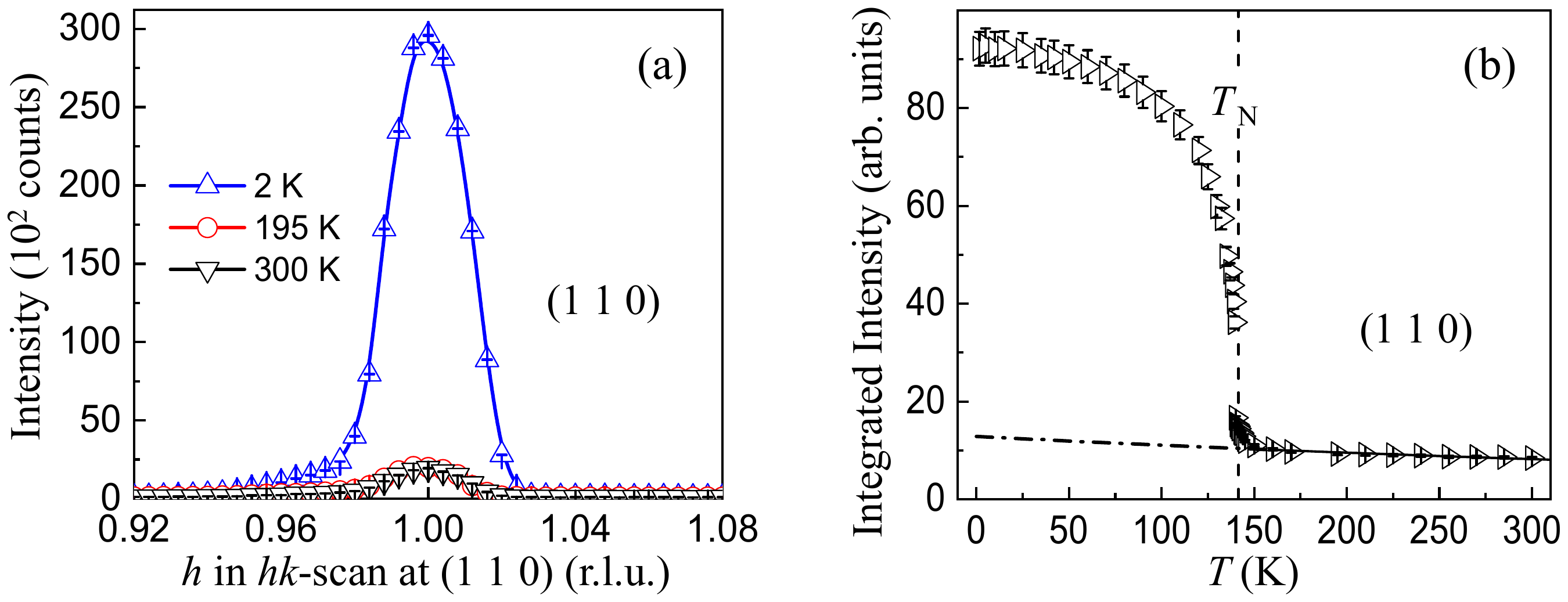}
\caption{
(a) Representative longitudinal scans of the magnetic Bragg (1 1 0) reflection at three temperatures of 2, 195, and 300 K, from the D23 (ILL, France) study on a YCrO$_3$ single crystal. The solid lines are guides to the eye.
(b) Corresponding temperature-dependent integrated intensities of the magnetic Bragg (1 1 0) reflection. $T_\textrm{N} = 141.5$(1) K labels the AFM transition temperature. The solid line was a fit to Eqs. (\ref{DW}) and (\ref{DW-2}) in the affiliated thermal regime. It was extrapolated to overall temperatures (dash-dotted line). The error bars in (a) and (b) are the standard deviations based on our measurements and fits.}
\label{Figure7}
\end{figure*}

\begin{table}[!t]
\renewcommand*{\thetable}{\Roman{table}}
\caption{Refined structural parameters of the pulverized YCrO$_3$ single crystal at 12, 145, and 300 K, including lattice constants, unit-cell volume, atomic positions, isotropic thermal parameters (\emph{B}), bond lengths, bond angles, and the distortion parameter $\Delta$ \cite{Zhu2019-2}. We listed the Wyckoff site of each ion and the goodness of fit. The numbers in parenthesis are the estimated standard deviations of the last significant digit.}
\label{Table2}
\begin{ruledtabular}
\begin{tabular} {llll}
\multicolumn {4}{c}{A pulverized YCrO$_3$ single crystal}                                                         \\
\multicolumn {4}{c}{(Orthorhombic, space group $Pnma$ (No. 62), $Z = 4$)}                                         \\
\hline
$T$ (K)                                       &12           &145           &300                                   \\
\hline
$a$ ({\AA})                                   &5.5189(1)    &5.5181(1)     &5.5198(1)                             \\
$b$ ({\AA})                                   &7.5205(1)    &7.5213(1)     &7.5297(1)                             \\
$c$ ({\AA})                                   &5.2323(1)    &5.2328(1)     &5.2392(1)                             \\
$\alpha (\beta, \gamma)$
$(^\circ)$                                    &90           &90            &90                                    \\
$V$ ({\AA}$^3$)                               &217.17(1)    &217.18(1)     &217.75(1)                             \\
\hline
Y(4\emph{c}) \emph{x}                         &0.0682(4)    &0.0672(1)     &0.0665(1)                             \\
Y(4\emph{c}) \emph{y}                         &0.25         &0.25          &0.25                                  \\
Y(4\emph{c}) \emph{z}                         &--0.0172(4)  &--0.0177(1)   &--0.0174(2)                           \\
Y(4\emph{c}) \emph{B} (\AA $^2$)              &0.2          &0.28(2)       &0.47(2)                               \\
\hline
Cr(4\emph{b}) $(x, y, z)$                     &(0, 0, 0.5)  &(0, 0, 0.5)   &(0, 0, 0.5)                           \\
Cr(4\emph{b}) \emph{B} (\AA $^2$)             &0.2          &0.29(3)       &0.39(3)                               \\
\hline
O1(4\emph{c}) \emph{x}                        &0.4643(5)    &0.4646(2)     &0.4647(2)                             \\
O1(4\emph{c}) \emph{y}                        &0.25         &0.25          &0.25                                  \\
O1(4\emph{c}) \emph{z}                        &0.1039(5)    &0.1052(2)     &0.1050(2)                             \\
O1(4\emph{c}) \emph{B} (\AA $^2$)             &0.2          &0.34(2)       &0.51(2)                               \\
\hline
O2(8\emph{d}) \emph{x}                        &0.3020(4)    &0.3020(1)     &0.3021(1)                             \\
O2(8\emph{d}) \emph{y}                        &0.0539(2)    &0.0538(1)     &0.0536(1)                             \\
O2(8\emph{d}) \emph{z}                        &--0.3065(4)  &--0.3067(1)   &--0.3066(1)                           \\
O2(8\emph{d}) \emph{B} (\AA $^2$)             &0.2          &0.34(2)       &0.51(2)                               \\
\hline
Y-O11 (\AA)                                   &2.237(3)     &2.231(1)      &2.233(1)                              \\
Y-O12 (\AA)                                   &2.276(4)     &2.286(1)      &2.290(1)                              \\
Y-O21 (\AA) ($\times 2$)                      &2.277(3)     &2.272(1)      &2.273(1)                              \\
Y-O22 (\AA) ($\times 2$)                      &2.476(3)     &2.479(1)      &2.485(1)                              \\
$<$Y-O$>$ (\AA)                               &2.337(1)     &2.337(1)      &2.340(1)                              \\
\hline
Cr-O1 (\AA) ($\times 2$)                      &1.967(1)     &1.969(1)      &1.971(1)                              \\
Cr-O21 (\AA) ($\times 2$)                     &1.983(2)     &1.983(1)      &1.984(1)                              \\
Cr-O22 (\AA) ($\times 2$)                     &1.992(2)     &1.991(1)      &1.993(1)                              \\
$<$Cr-O$>$ (\AA)                              &1.980(1)     &1.981(1)      &1.983(1)                              \\
\hline
$\angle$Cr-O1-Cr $(^\circ)$                   &145.81(3)    &145.49(1)     &145.53(1)                             \\
$\angle$Cr-O2-Cr $(^\circ)$                   &146.18(9)    &146.18(3)     &146.22(3)                             \\
\hline
$\Delta$(Y) $(\times 10^{-4})$                &18.197       &19.085        &19.789                                \\
$\Delta$(Cr) $(\times 10^{-4})$               &0.268        &0.215         &0.203                                 \\
$\Delta$(O1) $(\times 10^{-4})$               &47.429       &47.743        &47.930                                \\
$\Delta$(O2) $(\times 10^{-4})$               &90.063       &90.578        &91.684                                \\
\hline
$R_\textrm{p}$                                &6.40         &5.46          &6.03                                  \\
$R_\textrm{wp}$                               &9.33         &3.67          &3.74                                  \\
$R_\textrm{exp}$                              &7.33         &2.69          &2.83                                  \\
$\chi^2$                                      &1.62         &1.86          &1.75                                  \\
\end{tabular}
\end{ruledtabular}
\end{table}

\subsection{A. ICP-OES measurements}

By quantitative ICP-OES measurements, we determined the chemical compositions of the studied single crystal as Y$_{0.97(2)}$Cr$_{0.98(2)}$O$_{3.00(2)}$. This implies that the resultant single crystals of YCrO$_3$ compound by our FZ method are nearly stoichiometric within the experimental accuracy. Therefore, during analyzing magnetization and time-of-flight neutron-powder diffraction data, we kept the stoichiometry of the synthesized YCrO$_3$ samples being the ideal one (i.e., 1{:}1{:}3).

\subsection{B. Resistivity measurements}

We tried to measure possible resistivity in the YCrO$_3$ single crystals with a multimeter at room temperature. Unfortunately, it was beyond the maximum range (10$^6$ ohm) of the ohmmeter. In addition, our attempt to measure it by the standard four-probe method with our PPMS DynaCool system from 2 to 300 K was fruitless. Therefore, we conclude that the YCrO$_3$ compound is a robust insulator in our studied temperature range. A deeper understanding of the electronic states of conducting VO \cite{Greenwood1997} and insulating YCrO$_3$ compounds necessitates more experimental work and theoretical band structure calculations. Perhaps both samples are the only two pure 3$d^3$ compounds.

\subsection{C. Magnetization versus temperature}

Fig.~\ref{Figure2}(a) shows magnetization measurements of a small piece of randomly-orientated YCrO$_3$ single crystal. We transferred the unit of vertical axis into $\mu_\textrm{B}$ per Cr$^{3+}$ ion. There exists no obvious difference between ZFC and FC data. Upon cooling, ZFC and FC magnetization curves measured at 0.01 T show very small values down to temperature $\sim$141.5 K, e.g., ZFC magnetization $= 4.860(4) \times 10^{-7}$ and $7.310(6) \times 10^{-7} \mu_\textrm{B}/$Cr$^{3+}$ at 295 and 142.3 K, respectively. Around 141.5 K, they increase sharply by $\sim$38$\%$ in a small thermal range of $\sim$5 K, followed by a smooth increase down to 5 K. This resembles the characteristic feature of a reasonable canted antiferromagnet and rules out the possibility for a ferrimagnet. At 5 K, ZFC magnetization $= 4.530(4) \times 10^{-2} \mu_\textrm{B}/$Cr$^{3+}$.

We calculated the inverse magnetic susceptibility $\chi^{-1} = \mu_0H/M$ as shown in Fig.~\ref{Figure2}(b), where the nearly linear increase of $\chi^{-1}$ in the PM state at high temperatures obeys well the molar susceptibility via a CW law
\begin{eqnarray}
\chi(T) = \frac{C}{T - \theta_{\textrm{CW}}} = \frac{N_A \mu^2_{\textrm{eff}}}{3k_B(T - \theta_{\textrm{CW}})},
\label{CW}
\end{eqnarray}
where \emph{C} is the Curie constant, $\theta_{\textrm{CW}}$ is the PM Curie temperature, $N_A$ = 6.022 $\times$ 10$^{23}$ mol$^{-1}$ is the Avogadro's number, $\mu_{\textrm{eff}}$ = $g \mu_\textrm{B} \sqrt{J(J + 1})$ is the effective PM moment, and $k_B$ = 1.38062 $\times$ 10$^{-23}$ J/K is the Boltzmann constant. The fit by Eq.~(\ref{CW}) was shown as the dash-dotted line in Fig.~\ref{Figure2}(b). The fit parameters were listed in Table~\ref{Table1}.

Figures~\ref{Figure2}(a) and \ref{Figure2}(b) clearly indicate a sharp magnetic phase transition. We determined the magnetic phase transition temperature as $T_\textrm{N} =$ 141.5(1) K. The resultant PM CW temperature $\theta_{\textrm{CW}} =$ --433.2(6) K, indicating an existence of strong AFM correlations. We calculated the frustrating parameter \cite{Ramirez2001, Ramirez1994, Diep2004, Shores2005, HFLi2014, Li2015}, i.e., $f = \mid\theta_{\textrm{CW}}\mid/T_\textrm{N} =$ 3.061(5), which was consistent with our high-temperature magnetization study \cite{Zhu2019-2}. This value indicates that the low-temperature magnetic moments of Cr$^{3+}$ ions in YCrO$_3$ compound are frustrated by competitive spin exchanges. As listed in Table~\ref{Table1}, the extracted effective PM moment $\mu_{\textrm{eff{\_}meas}} =$ 3.95(2) $\mu_\textrm{B}$, a little bit larger than the calculated theoretical value $\mu_{\textrm{eff{\_}theo}} =$ 3.873 $\mu_\textrm{B}$, which was acceptable within the present experimental accuracy.

\subsection{D. Magnetization versus applied magnetic field}

Figure~\ref{Figure3} shows the measurement of magnetic hysteresis loop at 2 K. Figure~\ref{Figure3}(a) clearly exhibits the hysteresis loop whose shape is of a parallelogram. The loop locates in a magnetic field range of $\sim$--0.1 to 0.1 T with a coersive force of $\sim$0.05 T and a residual magnetism of $\sim$4.54 $\times 10^{-2}$ $\mu_\textrm{B}$/Cr$^{3+}$. These small values demonstrate that the YCrO$_3$ compound is a soft canted antiferromagnet at low temperatures. At 2 K and 7 T, the measured ZFC magnetization $M = 9.603 \times 10^{-2} \mu_\textrm{B}/$Cr$^{3+}$. From 0.1 to 7 T, the measured magnetization almost increases linearly with $\chi = M/\mu_0H =$ 7.24(1) $\times 10^{-3}$ $\mu_\textrm{B}$T$^{-1}$/Cr$^{3+}$. We therefore estimated that reaching a complete magnetic saturation state, an applied-magnetic field $\mu_0H \geq$ $\sim$408 T is required \cite{HFLi2016}.

\subsection{E. Heat capacity}

Figure~\ref{Figure4} shows the heat capacity measurements. At 0 T, with decreasing temperature, the measured heat capacity decreases until $T_\textrm{N} =$ 141.5 K, followed by an appearance of a $\lambda$-shape peak. Below this, heat capacity continues to decrease and gets flat below $\sim$24 K. The observation of the $\lambda$-shape peak indicates a phase transition. To reveal the nature of the phase transition, we measured heat capacity under an applied-magnetic field of 5 T. As shown in Fig.~\ref{Figure4}(a), at 5 T, it was noted that the intensity of the $\lambda$-shape peak was reduced, accompanied by a shift of the peak position from $\sim$138.8 K (0 T) to an elevated temperature $\sim$139.9 K (5 T). This is the characteristic feature of a canted antiferromagnet. Thus, the phase transition is magnetic rather than structural. We determined $T_\textrm{N}$ (5 T) {=} 144.5(1) K, $\sim$3 K higher than the $T_\textrm{N}$ at 0 T. Quantitative analysis of the relationship between values of $T_\textrm{N}$ and applied-magnetic-field strengths necessitates more measurements. It is pointed out that with the measurement of magnetization versus temperature, it is easy to determine the value of $T_\textrm{N}$. From heat capacity measurements, the phase transition temperature is at the temperature point at which a kink exists in the $C_\textrm{p} - T$ curve as marked in Fig.~\ref{Figure4}(a).

Albeit that the magnetization and heat capacity measurements show FM behaviors below $T_\textrm{N}$, the net magnetic interaction strength inside YCrO$_3$ compound is of strongly AFM because $\theta_{\textrm{CW}} =$ --433.2(6) K, indicating a complex low-temperature magnetic structure.

\subsection{F. Time of flight neutron powder diffraction}

To make the nature of the observed weak ferromagnetism clear and explore possible structural phase transitions in the YCrO$_3$ single crystal, we performed a time-of-flight neutron-powder diffraction study. The results were shown in Fig.~\ref{Figure5}. At the three temperatures as labeled, i.e., below (12 K), around (145 K), and above (300 K) the magnetic transition temperature ($\sim$141.5 K), all time-of-flight neutron-powder diffraction patterns were well indexed by an orthorhombic structure with the space group \emph{Pnma}. There was no detectable peak splitting or an appearance of satellite reflections. This indicates that no structural phase transition occurs in the YCrO$_3$ single crystal as a function of temperature in the studied thermal regime. This is in agreement with our heat capacity measurements. Based on the observed magnetic Bragg (0 1 1) and (1 1 0) peaks, as labeled in the right-bottom of Fig.~\ref{Figure5}, we established an AFM model with the propagation vector at \textbf{k} = (1 1 0) and the moment directions along the crystallographic \emph{c} axis. The extracted magnetic structure was schematically drawn in Fig.~\ref{Figure1}. It is pointed out that the magnetic (1 1 0) reflection is structurally forbidden by the space group \emph{Pnma}. We tried all possible canted AFM models, unfortunately, the corresponding FULLPROF refinements were not successful. The refined structural parameters were listed in Table~\ref{Table2}. It is pointed out that for the refinement of the data at 12 K, we constrained the isotropic thermal parameters (\emph{B}) of Y, Cr, O1, and O2 ions being the same as 0.2.

Within the present experimental accuracy, we can only determine a G-type AFM structure as shown in Fig.~\ref{Figure1}, where the nearest-neighbor Cr$^{3+}$ spins are aligned antiparallel. The directions of the AFM submoments are along with the crystallographic \emph{c} axis, i.e., the direction with the smallest lattice constant. It is hard to determine the possible canting angle. It was suggested that including a spin-orbital coupling on the quenched Cr$^{3+}$ ground state, an antisymmetric exchange interaction would cant the AFM moments along the crystallographic \emph{b} axis, i.e., the direction with the largest lattice constant, according to the Dzialoshinski-Moriya theory \cite{Cooke1974}. This leads to an appearance of the weak ferromagnetism.

Figure~\ref{Figure6} shows the refined moment size of Cr$^{3+}$ ions in the YCrO$_3$ single crystal, extracted from our time-of-flight neutron-powder diffraction study. As listed in Table~\ref{Table1}, the refined moment size at 12 K is 2.45(6) $\mu_\textrm{B}$, $\sim$82\% of the theoretical saturation moment (3 $\mu_\textrm{B}$), in agreement with our conclusion that there exists a magnetic frustration in YCrO$_3$ compound and the studies with x-ray magnetic circular dichroism and absorption spectroscopies \cite{Ahmed2019} where the computed values of spin and orbital moments are 2.38 $\mu_\textrm{B}$ and --0.094 $\mu_\textrm{B}$, respectively, and that the total magnetic moment has little contribution from the orbital component. As temperature increases, the refined moment size remains a plateau up to $\sim$50 K, followed by a gradual diminution with temperature upon warming until a radical disappearance around 141.5 K, the onset temperature of the AFM transition (Figs.~\ref{Figure2} and \ref{Figure6}). Above $T_\textrm{N}$ (0 T) {=} 141.5 K, intensities of the two magnetic Bragg reflections were undetectable.

\begin{figure}[!t]
\centering \includegraphics[width = 0.48\textwidth] {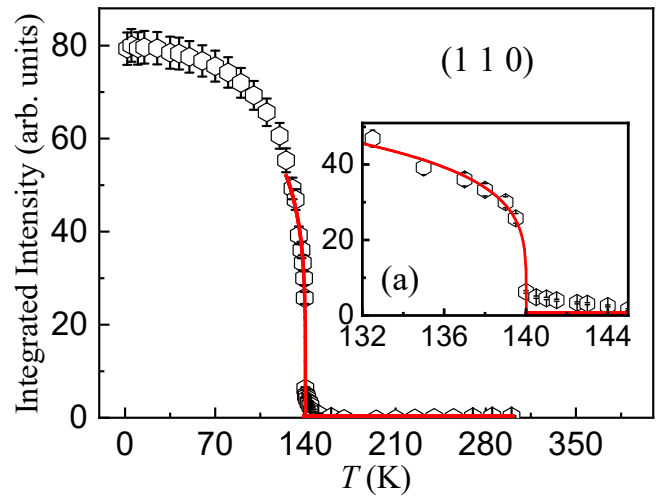}
\caption{
Subtracted integrated intensity from the pure magnetic contribution at the Bragg (1 1 0) peak position (void pentagons), to see detailed analysis in the text. Inset (a) shows the enlarged image around $T_\textrm{N}$ from 132--145 K. The solid line was a fit to the power-law Eq.~(\ref{DWIn-1}) in the affiliated thermal regime. The error bars are the propagated standard deviations based on our calculations.}
\label{Figure8}
\end{figure}

\subsection{G. Single crystal neutron diffraction}

An orientated YCrO$_3$ single crystal with the (\emph{H} \emph{K} 0) reciprocal-lattice vector in the scattering plane was used to perform a single-crystal neutron diffraction study on the D23 single-crystal diffractometer (ILL, France). Figure~\ref{Figure7}(a) shows some longitudinal scans of the magnetic Bragg (1 1 0) reflection. At 2 K (below $T_\textrm{N}$), we observed a very strong peak, indicating a formation of the AFM structure. At 195 and 300 K (above $T_\textrm{N}$), the intensity of the magnetic (1 1 0) peak decreased sharply but did not disappear. It is interesting that there still exists detectable intensity of the Bragg (1 1 0) peak above $T_\textrm{N}$. We ruled out the $\lambda/2$ contamination. As the foregoing remark, this reflection is forbidden by the space group \emph{Pnma}. Therefore, the existence of the Bragg (1 1 0) reflection above $T_\textrm{N}$ indicates that the actual crystalline structure of YCrO$_3$ compound may be lower than the orthorhombic structure with space group \emph{Pnma}. Our studies also demonstrate that the scattering ability of a single crystal is much higher than that of the corresponding pulverized powder sample.

At 195 and 300 K, the observed Bragg (1 1 0) reflection forbidden structurally by the \emph{Pnma} symmetry was treated to be from a pure nuclear contribution. The temperature variation of this contribution depends mainly on the thermal dynamic vibrations of related atoms, i.e., Debye-Waller (DW) factors. The falloff of the temperature-weakened intensity at a certain scattering vector \textbf{\emph{Q}} almost decays exponentially and can be estimated by
\begin{eqnarray}
I = I_0 e^{-2W(Q, T)},
\label{DW}
\end{eqnarray}
where the exponential part is the DW factor, and
\begin{eqnarray}
2W(Q,T) = \frac{\hbar^2Q^2}{2\textrm{M}} \int coth \left(\frac{\hbar\omega}{2k_BT}\right) \frac{Z(\omega)}{\omega} d\omega,
\label{DW-1}
\end{eqnarray}
where $\hbar$ = 1.054589 $\times$ 10$^{-34}$ J.s is the Planck constant divided by 2$\pi$, M is the atomic mass, and $Z(\omega)$ is the phonon density of states \cite{Stephen1984}. At high temperatures, it is given simply by \begin{eqnarray}
2\emph{W} = \frac{3\hbar^2Q^2}{\textrm{M}k_B\Theta^2_W}T,
\label{DW-2}
\end{eqnarray}
where $\Theta_W$ is the effective Debye temperature \cite{Kittel1987}. Therefore, we treated practically the DW factor with a linear response to temperature, as verified in previous studies \cite{Li2007_1, Li2007_2, Tari2003, Li2014Er}. We fit the integrated intensities of the Bragg (1 1 0) reflection above $T_\textrm{N}$ to Eqs. (\ref{DW}) and (\ref{DW-2}), shown as the solid line in Fig.~\ref{Figure7}(b), and extrapolated the fit to the entire temperature range (shown as the dash-dotted line).

\begin{figure}[!t]
\centering \includegraphics[width = 0.48\textwidth] {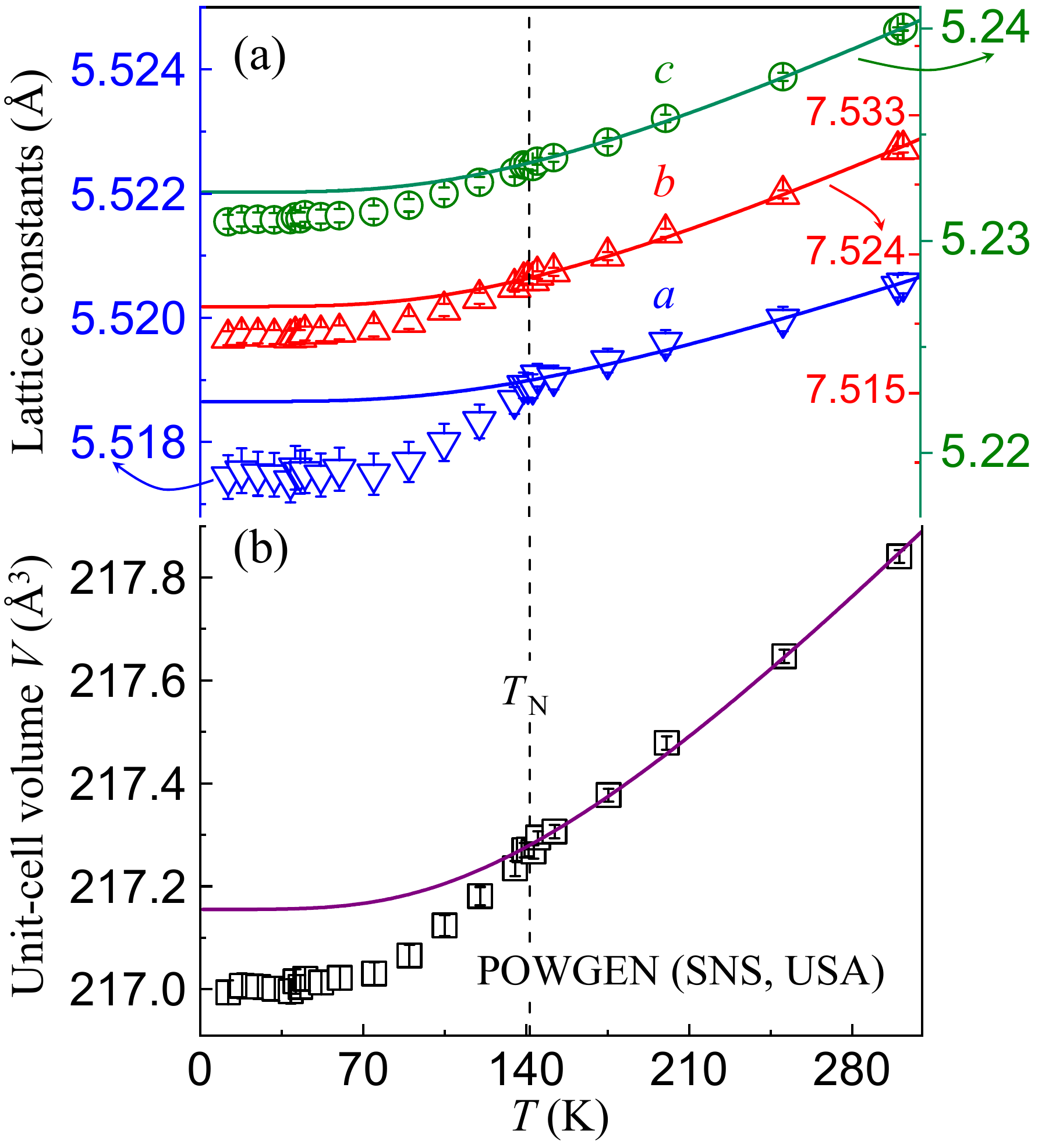}
\caption{
(a) Temperature-dependent lattice constants \emph{a}, \emph{b}, and \emph{c} of a pulverized YCrO$_3$ single crystal. (b) Corresponding anomalous unit-cell volume \emph{V} expansion with temperature. The solid lines in (a) and (b) are theoretical estimates of the variation of structural parameters using the Gr$\ddot{\textrm{u}}$neisen model with Debye temperature of $\theta_\textrm{D}$ = 580 K that is the same value as the one reported previously \cite{Zhu2019-2}. $T_\textrm{N}$ = 141.5(1) K labels the AFM transition temperature. The error bars in (a) and (b) are the standard deviations obtained from our FULLPROF refinements with the \emph{Pnma} structural symmetry.
}
\label{Figure9}
\end{figure}

\begin{figure*}[!t]
\centering \includegraphics[width = 0.78\textwidth] {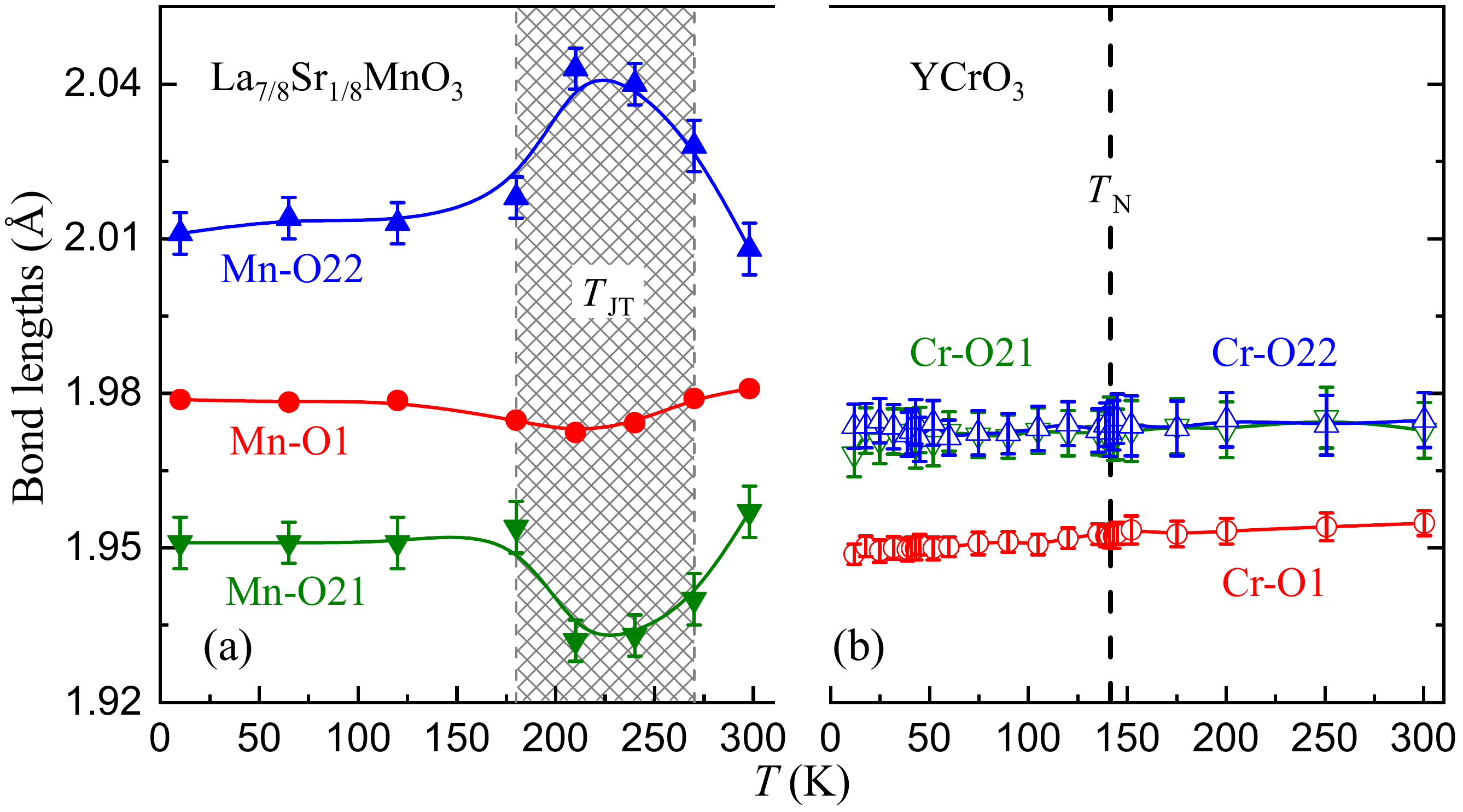}
\caption{
(a) Three bond lengths of Mn-O1, Mn-O21, and Mn-O22 in the La$_{\frac{7}{8}}$Sr$_{\frac{1}{8}}$MnO$_3$ single crystal \cite{Li2009} versus temperature. $T_\textrm{JT}$ $\approx$ 180--270 K denotes the regime of the Jahn-Teller effect. (b) Corresponding bond lengths in the YCrO$_3$ single crystal as a function of temperature from this study. $T_\textrm{N} = 141.5$(1) K labels the AFM transition temperature. The error bars in (a) and (b) were from our FULLPROF refinements.
}
\label{Figure10}
\end{figure*}

\begin{figure}[!t]
\centering \includegraphics[width = 0.48\textwidth] {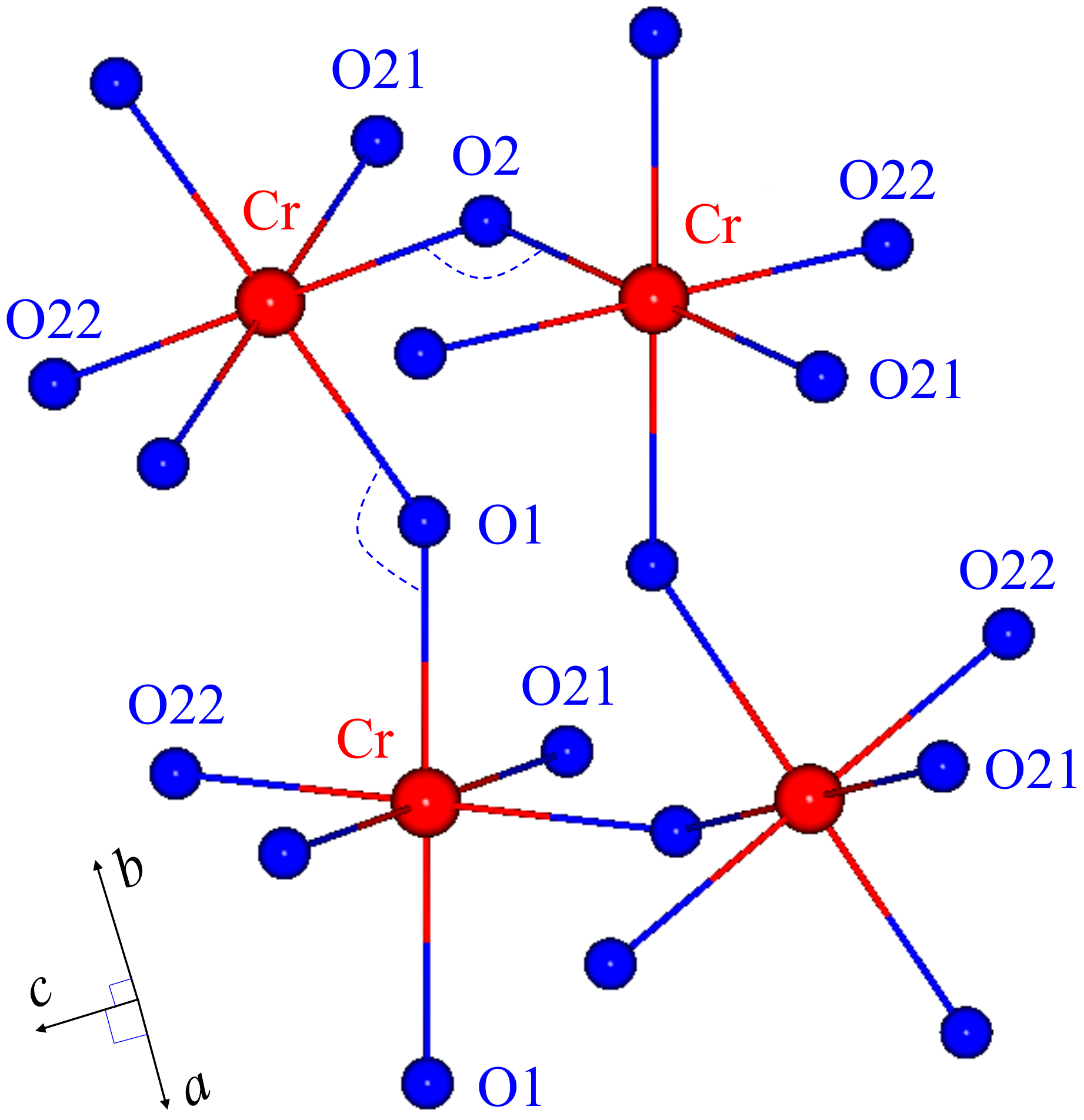}
\caption{
Schematic illustration of the three Cr-O bonds (Cr-O1, Cr-O21, and Cr-O22), as well as the two bond angles Cr-O-Cr (Cr-O1-Cr and Cr-O2-Cr) in the orthorhombic structure of a YCrO$_3$ single crystal. In this structural symmetry (with space group \emph{Pnma}), Cr ions in YCrO$_3$ compound have the same Wyckoff site, 4\emph{b} (0 0 0.5), as that of the Mn ions in La$_{\frac{7}{8}}$Sr$_{\frac{1}{8}}$MnO$_3$ compound \cite{Li2007_1, Li2007_2, Li2009, Li2008}.
}
\label{Figure11}
\end{figure}

\begin{figure}[!t]
\centering \includegraphics[width = 0.48\textwidth] {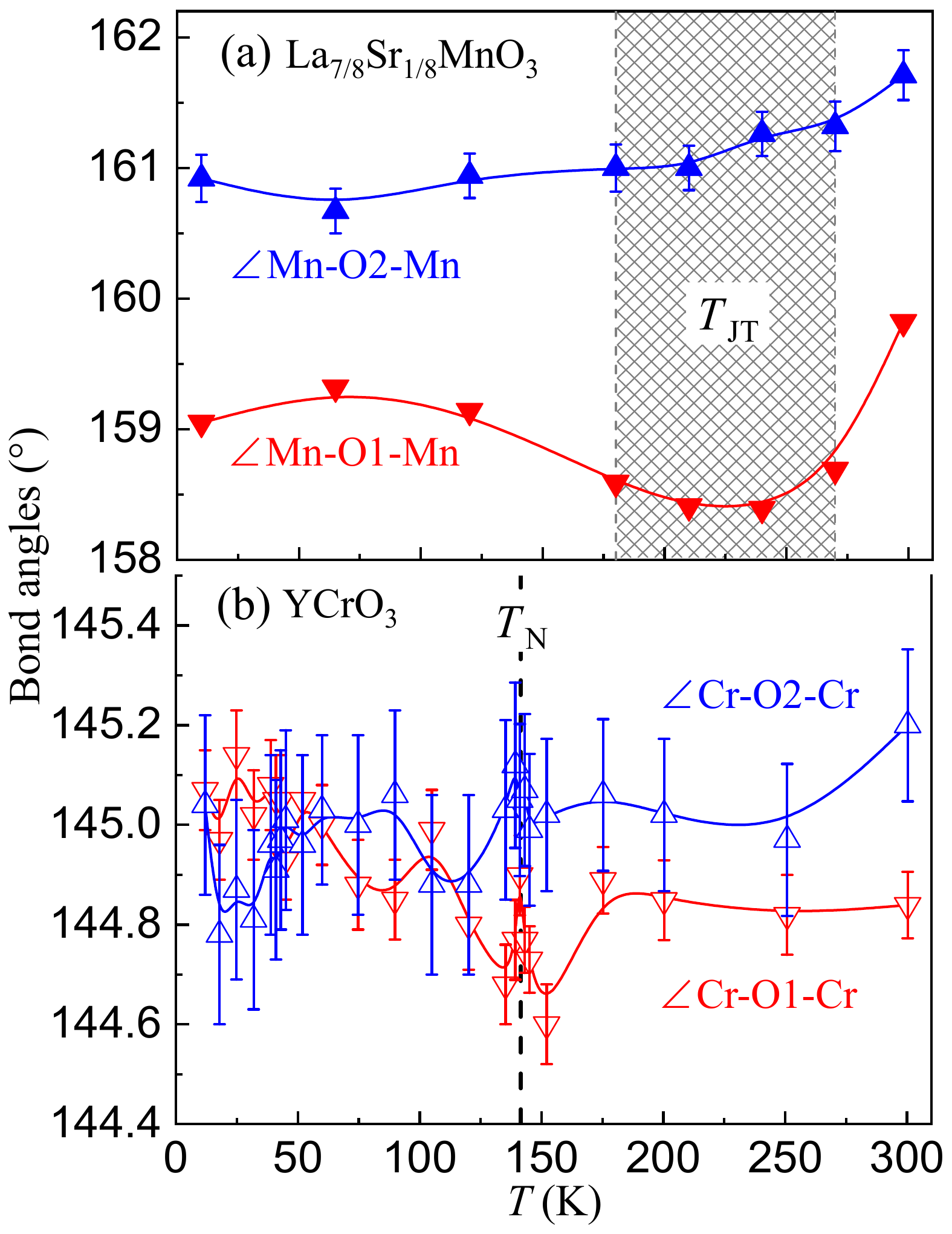}
\caption{
(a) Temperature-dependent bond angles of Mn-O1-Mn and Mn-O2-Mn in the La$_{\frac{7}{8}}$Sr$_{\frac{1}{8}}$MnO$_3$ single crystal \cite{Li2009}. $T_\textrm{JT}$ $\approx$ 180--270 K denotes the regime of the Jahn-Teller effect. (b) Temperature-dependent bond angles of Cr-O1-Cr and Cr-O2-Cr in the YCrO$_3$ single crystal from the present study. $T_\textrm{N} = 141.5$(1) K labels the AFM transition temperature. The error bars in (a) and (b) are the standard deviations from refinements. The solid lines in (a) and (b) are guides to the eye.
}
\label{Figure12}
\end{figure}

\begin{figure}[!t]
\centering \includegraphics[width = 0.48\textwidth] {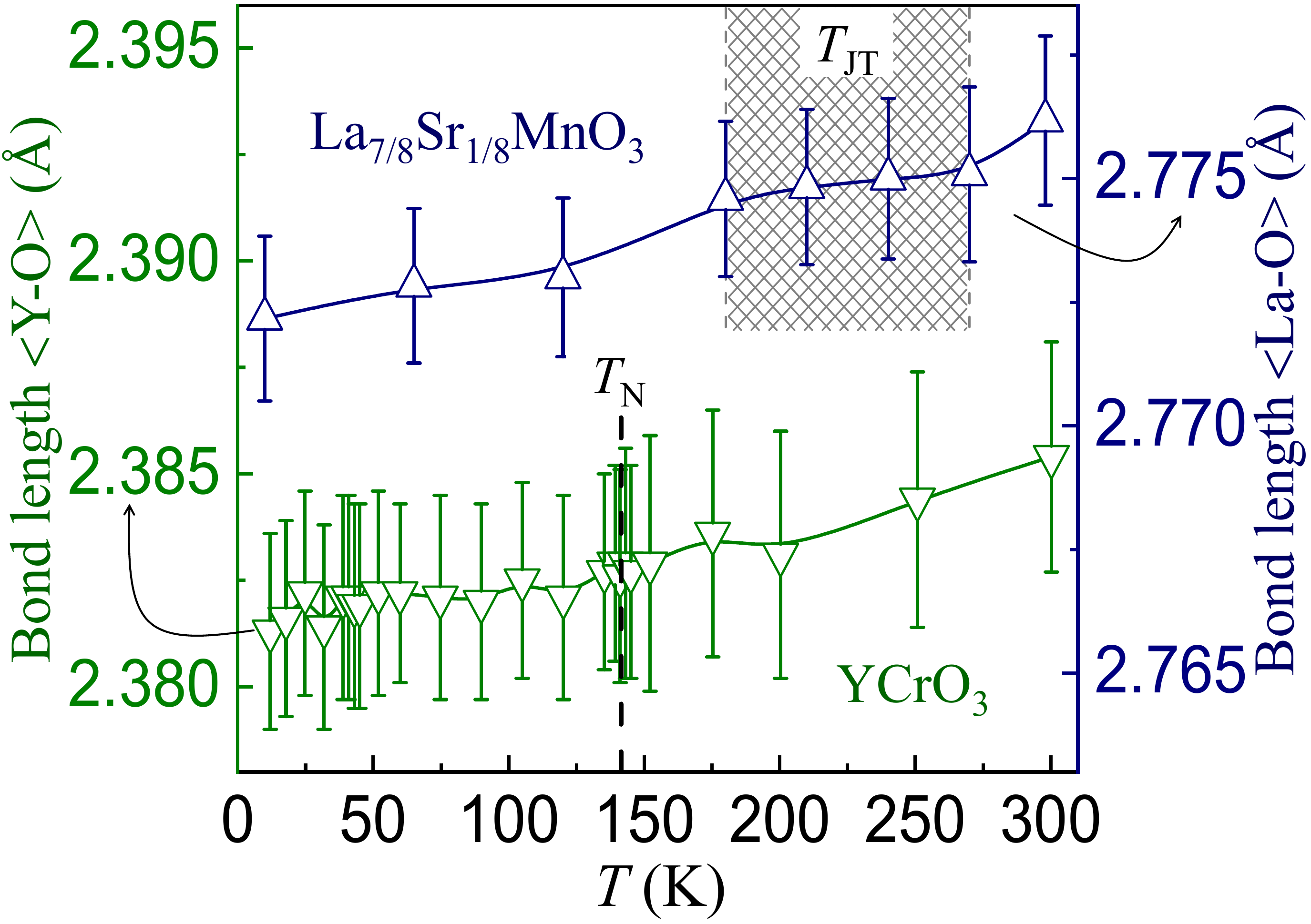}
\caption{
Comparison of the averaged bond lengths of Y-O in YCrO$_3$ (left, from the present study) and La-O in La$_{\frac{7}{8}}$Sr$_{\frac{1}{8}}$MnO$_3$ \cite{Li2009} (right) single crystals. $T_\textrm{JT}$ $\approx$ 180--270 K denotes the regime of the Jahn-Teller effect of La$_{\frac{7}{8}}$Sr$_{\frac{1}{8}}$MnO$_3$ compound. $T_\textrm{N} = 141.5$(1) K labels the AFM transition temperature of YCrO$_3$ compound. The error bars are the calculated standard deviations. The solid lines are guides to the eye. It is clear that the bond length of $\langle$Y-O$\rangle$ is shorter than that of the $\langle$La-O$\rangle$ bond beyond statistics.
}
\label{Figure13}
\end{figure}

Furthermore, we subtracted the corresponding nuclear component from the total scattered intensity at the Bragg (1 1 0) peak position to extract the pure magnetic contribution below $T_\textrm{N}$ as shown in Fig.~\ref{Figure8}. The resultant magnetic intensity above $T_\textrm{N}$ is approximately zero within accuracy, which in turn supports the above subtraction. The extracted integrated intensity (\emph{I}) of Bragg (1 1 0) reflection from the pure magnetic contribution can be fit to a power-law equation \cite{Collins1989, Payne1996}
\begin{eqnarray}
I(T) = I_0 \left( \frac{|T - T_{\textrm{N}}|}{T_{\textrm{N}}} \right)^{2\beta},
\label{DWIn-1}
\end{eqnarray}
where $T_{\textrm{N}}$ is the value of the N\'{e}el temperature, and $\beta$ is the critical exponent. Our fit with Eq.~(\ref{DWIn-1}) to the extracted data in a narrow thermal range from 125 to 140 K, shown as the solid line in Fig.~\ref{Figure8}, produces a N$\acute{\textrm{e}}$el temperature $T_{\textrm{N1}} =$ 140.0(1) K, and the critical exponent $2\beta =$ 0.215(6), indicating a second-order type phase transition and probably two-dimensional Ising-like spin interactions existing within the reciprocal (1 1 0) scattering plane \cite{Collins1989, Xiao2010}. The fitting procedure was as follow: First, we kept $I_0 =$ 80 and $T_\textrm{N} =$ 141.5 K and allowed $\beta$ to vary; finally, we fit all parameters together. For comparison, we further fit the data in three temperature ranges of 125--140 K, 130--140 K, and 135--140 K. No clear differences exist in the values of the refined $T_\textrm{N}$ and $\beta$, which validates our choice of the temperature range of 125--140 K for the final fitting.

As shown in Fig.~\ref{Figure8}(a), it is interesting to note that above $T_{\textrm{N1}}$, there exists weaker critical scattering over a range of temperature up to 145 K.

\subsection{H. Anisotropic magnetostriction effect}

The refined lattice parameters \emph{a}, \emph{b}, and \emph{c}, as well as the unit-cell volume \emph{V}, from our time-of-flight neutron-powder diffraction studies were shown in Fig.~\ref{Figure9} (void symbols). Upon cooling, the refined (Re) \emph{a}, \emph{b}, \emph{c}, and \emph{V} almost shrink linearly down to $T_\textrm{N}$ at which a cusp appears.

As in the foregoing discussions, the YCrO$_3$ compound is an insulator. We thus neglected the electronic contribution to the thermal expansion of its lattice configuration ($\varepsilon$). The temperature variation of the nonmagnetic contribution component is then mainly from phonons. This can approximately be estimated according to the Gr$\ddot{\textrm{u}}$neisen rules at zero pressure with a second-order fashion \cite{Li2012, Wallace1998, Vocadlo2002}
\setlength\arraycolsep{1.4pt} 
\begin{eqnarray}
\varepsilon(T) = \varepsilon_0 + \varepsilon_0\frac{U}{Q-BU},
\label{Gr1}
\end{eqnarray}
where $\varepsilon_0$ is the lattice parameter at zero Kelvin, and the internal energy \emph{U} can be calculated with the Debye approximations
\begin{eqnarray}
U(T) = 9Nk_BT\left(\frac{T}{\Theta_D}\right)^3 \int^{\frac{\Theta_D}{T}}_0 \frac{x^3}{e^x - 1}dx,
\label{Gr2}
\end{eqnarray}
where \emph{N} = 5 is the number of atoms per formula unit, and $\Theta_D$ is the Debye temperature. With Eqs. (\ref{Gr1}) and (\ref{Gr2}), we fit the lattice parameters of YCrO$_3$ compound in the PM state (above $T_\textrm{N}$ $\approx$ 141.5 K) and extrapolated the fits to overall temperatures as shown in Fig.~\ref{Figure9} (solid lines). For example, the fitting for the unit-cell volume \emph{V} results in $V_0$ $\approx$ 217.14 {\AA$^3$}, {\emph{Q}} {$\approx$} 7.57 $\times$ 10$^{-18}$ J, and \emph{B} $\approx$ --40.73. The different variations in \emph{a}, \emph{b}, and \emph{c} below $T_\textrm{N}$ in contrast to our theoretical estimates by the Gr$\ddot{\textrm{u}}$neisen (Gr) law \cite{Wallace1998, Vocadlo2002} (solid lines), e.g., $\frac{a^\textrm{12K}_\textrm{Re} - a^\textrm{12K}_\textrm{Gr}}{a^\textrm{12K}_\textrm{Gr}}$ $\approx$ --2.73 $\times$ 10$^{-4}$, $\frac{b^\textrm{12K}_\textrm{Re} - b^\textrm{12K}_\textrm{Gr}}{b^\textrm{12K}_\textrm{Gr}}$ $\approx$ --2.19 $\times$ 10$^{-4}$, and $\frac{c^\textrm{12K}_\textrm{Re} - c^\textrm{12K}_\textrm{Gr}}{c^\textrm{12K}_\textrm{Gr}}$ $\approx$ --2.67 $\times$ 10$^{-4}$, indicate an anisotropic magnetostriction effect and that magnetic anisotropy exists in YCrO$_3$ compound. Below $T_\textrm{N}$, the magnetically-driven additional decreases of \emph{a}, \emph{b}, and \emph{c} jointly result in an enhanced sample contraction upon cooling, e.g., $\frac{V^\textrm{12K}_\textrm{Re} - V^\textrm{12K}_\textrm{Gr}}{V^\textrm{12K}_\textrm{Gr}}$ $\approx$ --7.43 $\times$ 10$^{-4}$, signifying a magnetoelastic effect \cite{Li2011} and a localized nature of the $t_{2\textrm{g}}$ moments, opposite the case in the GdSi metallic compound \cite{Li2012}.

\subsection{I. Comparison between $t_{2\textrm{g}}$ YCrO$_3$ and $e_\textrm{g}$ La$_{7/8}$Sr$_{1/8}$MnO$_3$ compounds}

As shown in Fig.~\ref{Figure10}, it is of interest to compare the bond lengths of the 3\emph{d} Mn$^{3+}$/Mn$^{4+}$ ($e^1_\textrm{g}t^3_{2\textrm{g}}$/$e^0_\textrm{g}t^3_{2\textrm{g}}$) ions in La$_{\frac{7}{8}}$Sr$_{\frac{1}{8}}$MnO$_3$ compound \cite{Li2009, Kotani2017} with those of the Cr$^{3+}$ ($e^0_\textrm{g}t^3_{2\textrm{g}}$) ions in YCrO$_3$ compound. The structural parameters of La$_{\frac{7}{8}}$Sr$_{\frac{1}{8}}$MnO$_3$ compound \cite{Li2009} were from a neutron-powder diffraction study on samples pulverized from a single crystal. This is thus comparable to the results from the present study. Both bond lengths of Mn-O21 and Mn-O22 [Fig.~\ref{Figure10}(a)] and the bond angle of Mn-O1-Mn [Fig.~\ref{Figure12}(a)] respond readily to the Jahn-Teller effect that occurs in La$_{\frac{7}{8}}$Sr$_{\frac{1}{8}}$MnO$_3$ compound within a temperature range of $\sim$180--270 K, whereas, those in YCrO$_3$ compound keep nearly constants, and both Cr-O bond lengths [Fig.~\ref{Figure10}(b)] and Cr-O-Cr bond angles [Fig.~\ref{Figure12}(b)] exhibit no response to the AFM transition, consistent with the fact that Cr$^{3+}$ ions don't have the Jahn-Teller effect. The values of Cr-O1 bond lengths are similar to those of the Mn-O21, and Cr-O21 and Cr-O22 to Mn-O1 (Fig.~\ref{Figure10}).

As shown in Figs.~\ref{Figure11} and \ref{Figure12}, in contrast to the bond angle of Mn-O-Mn, $\angle$Cr-O-Cr decreases hugely by $\sim$15$^\circ$, which in our opinion corresponds intimately to a possible lowering of the crystalline symmetry in the YCrO$_3$ single crystal. The relatively shorter bond length of $\langle$Y-O$\rangle$ (Fig.~\ref{Figure13}) introduces an immense mismatch between Y$^{3+}$ and Cr$^{3+}$ sites, leading to a huge chemical pressure and driving the subsequent rotating and tilting of the CrO$_6$ octahedra.

\section{IV. Conclusions}

In summary, we have grown a nearly stoichiometric Y$_{0.97(2)}$Cr$_{0.98(2)}$O$_{3.00(2)}$ single crystal by a laser diode FZ furnace. There are three electrons locating on the 3\emph{d} $t_{\textrm{2g}}$ orbitals of Cr$^{3+}$ ions, therefore, the YCrO$_3$ compound is a robust insulator. Although the measurements of the applied-magnetic-field dependent heat capacity as well as the magnetization versus temperature and applied-magnetic field show the character of a very soft ferromagnet with a coersive force of $\sim$0.05 T, the extracted PM CW temperature, $\theta_{\textrm{CW}} =$ --433.2(6) K, by the fit with a CW law is strongly negative with the frustrating parameter $f = \mid\theta_{\textrm{CW}}\mid/T_\textrm{N} =$ 3.061(5), and the measured magnetization at 2 K and 7 T is only $\sim$3.2\% of the theoretical saturation moment. These indicate that the spin moments of Cr$^{3+}$ ions in YCrO$_3$ compound are magnetically frustrated. The consistency between the effective PM moment, $\mu_{\textrm{eff{\_}meas}} =$ 3.95(2) $\mu_\textrm{B}$, and the theoretically-calculated value, $\mu_{\textrm{eff{\_}theo}} =$ 3.873 $\mu_\textrm{B}$, validates our results concluded in the framework of the CW-law fitting. By magnetization measurements, we determined the magnetic phase transition temperature as $T_\textrm{N} =$ 141.5(1) K at an applied-magnetic field of 0.01 T. This is in agreement with our neutron-powder and single-crystal diffraction studies. The magnetic transition temperature was pushed upward to $T_\textrm{N}$(5T) = 144.5(1) K at 5 T, increased by $\sim$3 K.

With our neutron-powder diffraction study, we have established an AFM structure with the propagation vector at \textbf{k} = (1 1 0) and the same unit cell as that of the crystalline structure (with space group $Pnma$). The direction of the Cr$^{3+}$ spin moments is along the crystallographic \emph{c} axis. The refined moment size is 2.45(6) $\mu_\textrm{B}$ at 12 K, $\sim$82\% of the theoretical saturation value 3 $\mu_\textrm{B}$. This is consistent with the fact that a magnetic frustration exists in YCrO$_3$ compound. By fitting integrated intensities of the magnetic Bragg (1 1 0) reflection extracted from the pure magnetic contribution with a power law, we found that the Cr$^{3+}$ spin interactions were probably two-dimensional Ising like within the reciprocal (1 1 0) scattering plane. Above $T_\textrm{N} =$ 141.5(1) K, the refined lattice constants \emph{a}, \emph{b}, and \emph{c}, as well as the unit-cell volume \emph{V}, agree well with the Gr$\ddot{\textrm{u}}$neisen rules at zero pressure with a second-order fashion. By comparison, below $T_\textrm{N}$, the lattice configuration (\emph{a}, \emph{b}, \emph{c}, and \emph{V}) deviates largely downward from the Gr$\ddot{\textrm{u}}$neisen law, displaying an anisotropic magnetostriction effect along the crystallographic \emph{a}, \emph{b}, and \emph{c} axes and a magnetoelastic effect with the unit-cell volume \emph{V}. Especially, upon cooling, the sample contraction is enhanced below $T_\textrm{N}$.

In the whole studied temperature range of 12--302 K, we did not find any crystalline structural phase transition with the neutron-powder diffraction study, whereas by our single-crystal neutron diffraction study, we observed clearly the existence of the Bragg (1 1 0) peak above the magnetic phase transition temperature 141.5(1) K. This peak persists up to 300 K and is forbidden by the crystalline orthorhombic structure (with space group $Pnma$). This implies that the actual crystalline structure of YCrO$_3$ compound is probably lower than the present one. To figure out the reasons for a possible lowering of the structural symmetry in the YCrO$_3$ single crystal, we compared the $t_{2\textrm{g}}$ YCrO$_3$ and the $e_\textrm{g}$ La$_{7/8}$Sr$_{1/8}$MnO$_3$ single crystals. It is pointed out that with a limited number of the observed magnetic Bragg peaks, it is hard to determine the canting degree of the AFM Cr$^{3+}$ spins. This can be addressed by a further time-of-flight single-crystal neutron-diffraction study. To determine the actual crystalline structure is not an easy job, but it would shed light on the dielectric anomaly of YCrO$_3$ compound.

\section{Acknowledgements}

T.L. acknowledges the National Natural Science Foundation of China (Grant No. 11604214), the Foundation of Department of Education of Guangdong Province (Grant No. 2018KTSCX223), and the Foundation of Department of Science and Technology of Guangdong Province (Grant No. 2020A1515010814).
J.M. acknowledges the National Science Foundation of China (Grant No. 11974042).
D.O. acknowledges financial support from the Science and Technology Development Fund, Macao SAR (File No. 0029/2018/A1).
H.-F.L. acknowledges the University of Macau (Files No. SRG2016{--}00091{--}FST and No. MYRG2020--00278--IAPME), the Science and Technology Development Fund, Macao SAR (Files No. 063/2016/A2, No. 064/2016/A2, No. 028/2017/A1, and No. 0051/2019/AFJ), and the Guangdong--Hong Kong--Macao Joint Laboratory for Neutron Scattering Science and Technology.
This research used resources at the Spallation Neutron Source, a Department of Energy Office of Science User Facility operated by Oak Ridge National Laboratory.


\end{document}